\documentclass[pra,superscriptaddress,twocolumn,amsmath,amssymb]{revtex4}
\usepackage{amsfonts}
\usepackage{amssymb}
\usepackage{mathtools}
\usepackage{graphicx}
\usepackage{subfigure}
\usepackage{mathrsfs}
\usepackage{color}
\usepackage[normalem]{ulem}%avoids the unnecessary underlining of references
\usepackage{natbib}
\usepackage{bbold}
\usepackage{placeins}
\usepackage{braket}
\usepackage{soul,xcolor}
\usepackage{epstopdf}
\usepackage{tabu}
\usepackage{array}
\usepackage{bm}
\usepackage{upgreek}
\usepackage{multirow}
\usepackage[utf8]{inputenc}
\usepackage[colorlinks = truelinkcolor = blue, urlcolor  = blue, citecolor = blue, anchorcolor = blue]{hyperref}
\hypersetup{
	colorlinks   = true, %Colours links instead of ugly boxes
	urlcolor     = blue, %Colour for external hyperlinks
	linkcolor    = blue, %Colour of internal links
	citecolor   = blue %Colour of citations
}
\makeatletter

\begin{document} 
	
	\title{Violation of Leggett-Garg type inequalities in a driven two level atom interacting with a squeezed thermal reservoir}
	
	\author{Javid Naikoo}
\email{naikoo.1@iitj.ac.in}
\affiliation{Indian Institute of Technology Jodhpur, Jodhpur 342011, India}

\author{Subhashish Banerjee}
\email{subhashish@iitj.ac.in}
\affiliation{Indian Institute of Technology Jodhpur, Jodhpur 342011, India}

	\author{Arun M. Jayannavar}
	\email{jayan@iopb.res.in}
	\affiliation{Institute of Physics, Bhubaneswar, India}

	\begin{abstract}
		\noindent 
		\textbf{Abstract}
	     The violation of Leggett-Garg type inequalities (LGtIs) is studied on a two level atom, driven by an external field in the presence of a squeezed thermal reservoir. The violations are observed in the  underdamped regime where the spontaneous transition rate is much smaller compared to the Rabi frequency. Increase in thermal effects is found to decrease the extent of violation as well as the time over which the violation lasts. With  increase in the value squeezing parameter the extent of violation of LGtIs is seen to reduce. The violation of LGtIs is favored by increase in the driving frequency.  Further, the interplay of the degree of violation and strength of the  measurements is studied. It is found that  the maximum violation occurs for ideal projective measurements.
	\end{abstract}
	
	\maketitle

	\section{Introduction}
	  Quantum mechanics  is so far the most  elegant interpretation of nature whose predictions have been verified in various experiments.  Central to quantum mechanics are the notions like  coherence and entanglement  arising from the superposition principle \cite{schrodinger1935,EPR}. Various approaches have been developed for quantification of quantumness leading to  computable measures of nonclassicality  \cite{Bell1964,horodecki2009quantum}. Another way of assessing the quantum coherent evolution is via inequalities based on  the time correlation functions, know as Leggett-Garg inequlities (LGIs).
	  
	  The LGIs have been developed to test the quantum coherence at macroscopic level \cite{leggett1985quantum,emary2013leggett}. These inequalities are based on the assumptions of \textit{macrorealism} and \textit{noninvasive} measurability. The former assigns well defined macroscopically distinct states to an observable irrespective of the observation, while  the later ensures that the post measurement dynamics is unaffected by the act of measurement. A quantum mechanical system does not obey these assumptions. The superposition principle violates  \textit{macrorealism} and the collapse postulate nullifies the possibility of a  noninvasive measurement.  
	  
	  The verification of  LGIs involve a single system being measured at different times unlike Bell inequality which involves multiple parties spatially separated from each other \cite{Inranil}.  
	 The simplest Leggett-Garg inequality is the one corresponding to three time measurements made at times $t_0$, $t_1$ and $t_2$ such that $t_0<t_1<t_2$. For a dichotomic operator $\hat{\mathcal{M}}(t)$, we define the two time correlation function $C(t_i, t_j) = \langle \hat{\mathcal{M}}(t_i) \hat{\mathcal{M}}(t_j) \rangle = \operatorname{Tr}[\rho \hat{\mathcal{M}}(t_i) \hat{\mathcal{M}}(t_j)]$. For the three time measurement case, we define the following combination of the two time correlation functions $K_3 = C(t_0,t_1) + C(t_1,t_2) - C(t_0,t_2)$, such that the simplest LGI reads
	 \begin{equation}
	 -3 \le K_3 \le 1.
	 \end{equation}
	 A violation of either lower or the upper bound is a signature of the ``quantumness" of the system. The two time correlation function can be evaluated as follows,
	 \begin{equation}\label{eq:TTC}
	 C(t_i, t_j) = \sum\limits_{m,n = \pm} m n \operatorname{Tr}\Big[\Pi^m \mathcal{E}_{t_j \leftarrow t_i} \big[ \Pi^n \rho(t_i) \Pi^n \big] \Big].
	 \end{equation}
	 Here, $\mathcal{E}_{t_b \leftarrow t_a}$ is the map governing the time evolution of the state, i.e., $\rho(t_b) = \mathcal{E}_{t_b \leftarrow t_a} [\rho(t_a)]$. The LGIs have been part of many theoretic \cite{barbieri2009multiple,avis2010leggett,lambert2010distinguishing,lambert2011macrorealism,montina2012dynamics,kofler2013condition,budroni2013bounding,Swati2017,javid2018Studyoftemporal,javid2019LGI3flavor,javid2018LGINM,naikoo2018entropic} and experimental \cite{palacios2010experimental,groen2013partial,goggin2011violation,dressel2011experimental,suzuki2012violation,athalye2011investigation,souza2011scattering,katiyar2013violation} studies. \par
	 In this work, we deviate from the original formulation of LGI and study instead a variant form of it, known as Leggett-Garg type inequalities (LGtIs) introduced in  \cite{Huelga1995,Huelga1996,Waldherr2011} and  experimentally verified in \cite{xu2011experimental,ZhouLGtIexp2015}. These inequalities were derived to avoid the requirement of noninvasive measurements at intermediate times. This feature makes them more suitable for the experimental verification as compared to LGIs. The assumption of NIM  is replaced by a weaker condition known as \textit{stationarity}. This asserts that the conditional probability $p(\phi, t_j| \psi, t_i)$ that the system is found in state $\phi$ at time $t_j$ given that it was in state $\psi$ at time $t_i$ is a function of the time difference $(t_j-t_i)$. Invoking stationarity leads to the following form of LGtIs
	 \begin{equation}\label{eq:LGtI}
	 K_{\pm} = \pm 2 C(t_0,t) - C(t_0,2t) \le 1.
	 \end{equation}
	 Here, $t = t_2-t_1 = t_1-t_0$, is the time between two successive measurements. From here on, we will call $K_{\pm}$ as  LG parameter.	 Though the assumption of stationarity helps to put the inequalities into easily testable forms, it reduces the class of macrorealist theories which are put to the test \cite{Huelga1995}. The stationarity condition holds provided  the system can be prepared in a well-defined state and the system evolves under Markovian dynamics. These conditions are satisfied in the model considered in this work. Therefore, for a suitable experimental setup, inequalities (\ref{eq:LGtI}) provide a tool to quantitatively probe the coherence effects in this system.
	
	 Here we study the violation of LGtIs in a driven two-level atom  interacting with  a squeezed thermal reservoir. The paper is organized as follows. In Sec. (\ref{Model}), we discuss in detail the model considered. Section (\ref{LGI}) is devoted to the description of LGtIs in the context of the model considered. The results and their discussion are given in Sec. (\ref{R&D}). We conclude in Sec. (\ref{Conclusion}).
	
	 \section{Model: A driven two level system}\label{Model}
\begin{figure}
	 	\centering
	 	\includegraphics[width=70mm]{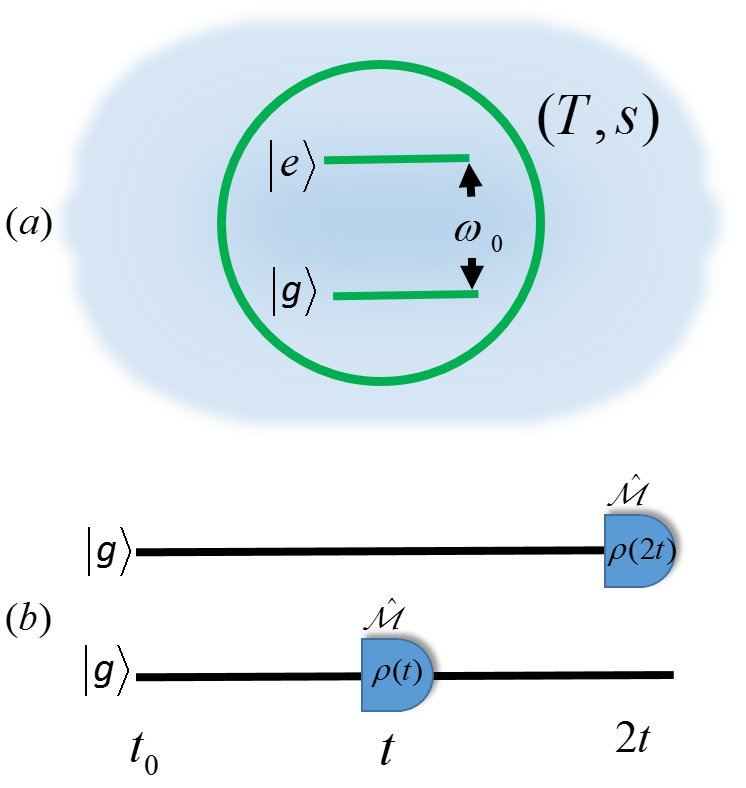}
	 	\caption{(color online).   Schematic diagram for (a) Two level atom interacting with a squeezed thermal bath at temperature $T$ with squeezing parameter $s$. The transition frequency between the two levels is $\omega_0$. (b)  Testing the LGtIs using the statistics of two experiments, with the same preparation state, $\ket{g}$, at time $t_0=0$. The dichotomic observable $\hat{\mathcal{M}}= \ket{g}\bra{g} - \ket{e}\bra{e}$ would lead to $+1$ if the atom is found in ground state and $-1$ otherwise.  For example, at $t_0$, we have $\langle \hat{\mathcal{M}}\rangle= +1$.}
	 	\label{fig:ModelExp}
	 \end{figure}

	 Here, we sketch the essential details of a driven two-level system in contact with a squeezed thermal bath  \cite{PhysRevA.77.012318,banerjee2008geometric,BP,banerjee2018open,Omkar2013}. 	 The model consists of a two level system whose Hilbert space is spanned by two states, the ground state $\ket{g}$ and the excited state $\ket{e}$, Fig.(\ref{fig:ModelExp}). The description of such a system is analogous to that of a spin -$\frac{1}{2}$ system. The Pauli operators in terms of these basis vectors are  $\sigma_1 = |e\rangle \langle g| + |g\rangle \langle e|$, $\sigma_2 = -i|e\rangle \langle g| + i |g \rangle \langle e|$ and $\sigma_3 = |e\rangle \langle e| - |g\rangle \langle g|$, and satisfy the usual commutation  $[\sigma_i, \sigma_j] = 2 i \epsilon _{ijk} \sigma_k$ and the anticommutation $\{\sigma_i, \sigma_j\} = 2\delta_{ij}$. The raising and lowering operators can be defined as 
	 \begin{align}
	 \sigma_+ &= |e\rangle \langle g| = \frac{1}{2} (\sigma_1 + i \sigma_2),\nonumber\\
	  \sigma_- &= |g\rangle \langle e|  =\frac{1}{2}(\sigma_1 - i \sigma_2).
	 \end{align}
	 With this setting, we can define the system Hamiltonian $H_S$ to be diagonal in basis $\{\ket{e},\ket{g}\}$. With $\omega_0$ denoting the transition frequency between the two levels (setting $\hbar = 1$), we have
	 \begin{equation}
	 H_S = \frac{1}{2} \omega_0 \sigma_3.
	 \end{equation}
	 A detailed account of two level systems and their application can be found in \cite{feynman2011feynman}.\par
	 We now consider the case when a two level atomic transition $\ket{e} \leftrightarrow \ket{g}$ is driven by an external source. The source is assumed to be a coherent single mode field on resonance. Under dipole approximation, the  Hamiltonian (in the interaction picture) is given by $H_L = - \vec{E}_L(t). \vec{D}(t)$. Here, $\vec{E}_L(t) = \vec{\epsilon} e^{-i \omega_0 t} + \vec{\epsilon}^* e^{+i \omega_0 t} $ is the electric field strength of the driving mode. Also, $\vec{D}(t) = \vec{d} \sigma_{-} e^{-i\omega_0 t} + \vec{d}^* \sigma_{+} e^{+i\omega_0 t}$ is the atomic dipole operator in the interaction picture and $\vec{d}= \langle g|\vec{D}|e\rangle$ is the transition matrix element of the dipole operator. The atom-field interaction can be written in the rotating wave approximation as follows,
	 \begin{equation}
	 H_L= - \frac{\Omega}{2} (\sigma_+ + \sigma_-).
	 \end{equation}
	 Here, $\Omega = 2 \vec{\epsilon}.\vec{d}^*$, is referred to as the Rabi frequency. Now coupling the system to  a thermal reservoir leads to the quantum master equation
	 \begin{align}\label{eq:master}
	 \frac{d \rho(t)}{d t} &= \frac{i \Omega}{2} \big[\sigma_{+} + \sigma_{-}, \rho(t) \big] \nonumber \\&+ \gamma_0 n \bigg( \sigma_{+} \rho(t) \sigma_{-} - \frac{1}{2} \sigma_{-} \sigma_{+} \rho(t)  - \frac{1}{2} \rho(t) \sigma_{-} \sigma_{+}\bigg) \nonumber \\&+ \gamma_0 (n+1) \bigg( \sigma_{-} \rho(t) \sigma_{+} - \frac{1}{2} \sigma_{+} \sigma_{-} \rho(t)  - \frac{1}{2} \rho(t) \sigma_{+} \sigma_{-}\bigg) \nonumber \\&- \gamma_0 M \sigma_{+} \rho(t) \sigma_{+} - \gamma_0 M^* \sigma_{-} \rho(t) \sigma_{-}.
	 \end{align}
	 Here, $\gamma = \gamma_0 (2n + 1)$ is the total transition rate with  $\gamma_0$ being the spontaneous emission rate. Further, 
	 \begin{align}\label{eq:NandM}
	 n &= n_{th} (\cosh^2(s) + \sinh^2(s) ) + \sinh^2(s), \nonumber \\
	 {\rm and} ~~M &= -\cosh(s) \sinh(s) e^{i \theta} (2 n_{th} + 1).
	 \end{align}
	  where $s$ and $\theta$ are the squeezing parameters and $n_{th} = 1/(\exp[\beta \omega_0] - 1)$ is the Plank distribution at  transition frequency. In what follows, we will set $\theta = 0$ for the purpose of calculations. \par
	 In order to solve Eq. (\ref{eq:master}), we write the density matrix as 
	  \begin{align}
	  \rho(t) &= \frac{1}{2} (\mathbf{I} + \vec{v}(t).\vec{\sigma}) 
	            = \begin{pmatrix}
	                  \frac{1}{2}(1 + \langle \sigma_3 \rangle )     & \langle \sigma_{-} \rangle\\
	                  \langle \sigma_{+} \rangle          &         \frac{1}{2}(1 - \langle \sigma_3 \rangle )
	            \end{pmatrix},
	  \end{align}
	  with $\vec{v}(t)=\langle \vec{\sigma}(t) \rangle = \operatorname{Tr}[\vec{\sigma} \rho(t)]$, is known as the Bloch vector.
	With this notation,  the master equation, Eq. (\ref{eq:master}), becomes
	 \begin{equation}\label{eq:BlochEq}
	 \frac{d}{d t} \langle \vec{\sigma}(t) \rangle = \mathcal{G} \langle \vec{\sigma}(t) \rangle + \vec{m}.
	 \end{equation}
	 Here, 
	 \begin{equation}
	 \mathcal{G} =  \begin{pmatrix}
	 -\frac{\gamma}{2} - \gamma_0 M   & 0  &  0 \\\\
	 0  &    -\frac{\gamma}{2} + \gamma_0 M  &  \Omega\\\\
	 0                                &             - \Omega                         &        - \gamma
	 \end{pmatrix},
	 \end{equation}
	 and $\vec{m} =[0~~ 0~~ -\gamma_0]^T$, $T$ being the transpose operation.
	
	 The differential equation (\ref{eq:BlochEq}) has the stationary solution given by 
	 \begin{align}
	  \langle \sigma_3 \rangle_s &= -\frac{\gamma_0 (\gamma -  2 \gamma_0 M)}{\gamma^2 - 2 \gamma \gamma_0 M + 2 \Omega^2},\nonumber \\
	  \langle \sigma_+ \rangle_s &= -\frac{ i \gamma_0 \Omega }{\gamma^2 - 2 \gamma \gamma_0 M + 2 \Omega^2}.
	 \end{align}
	  \begin{figure}
	 	\includegraphics[width=80mm]{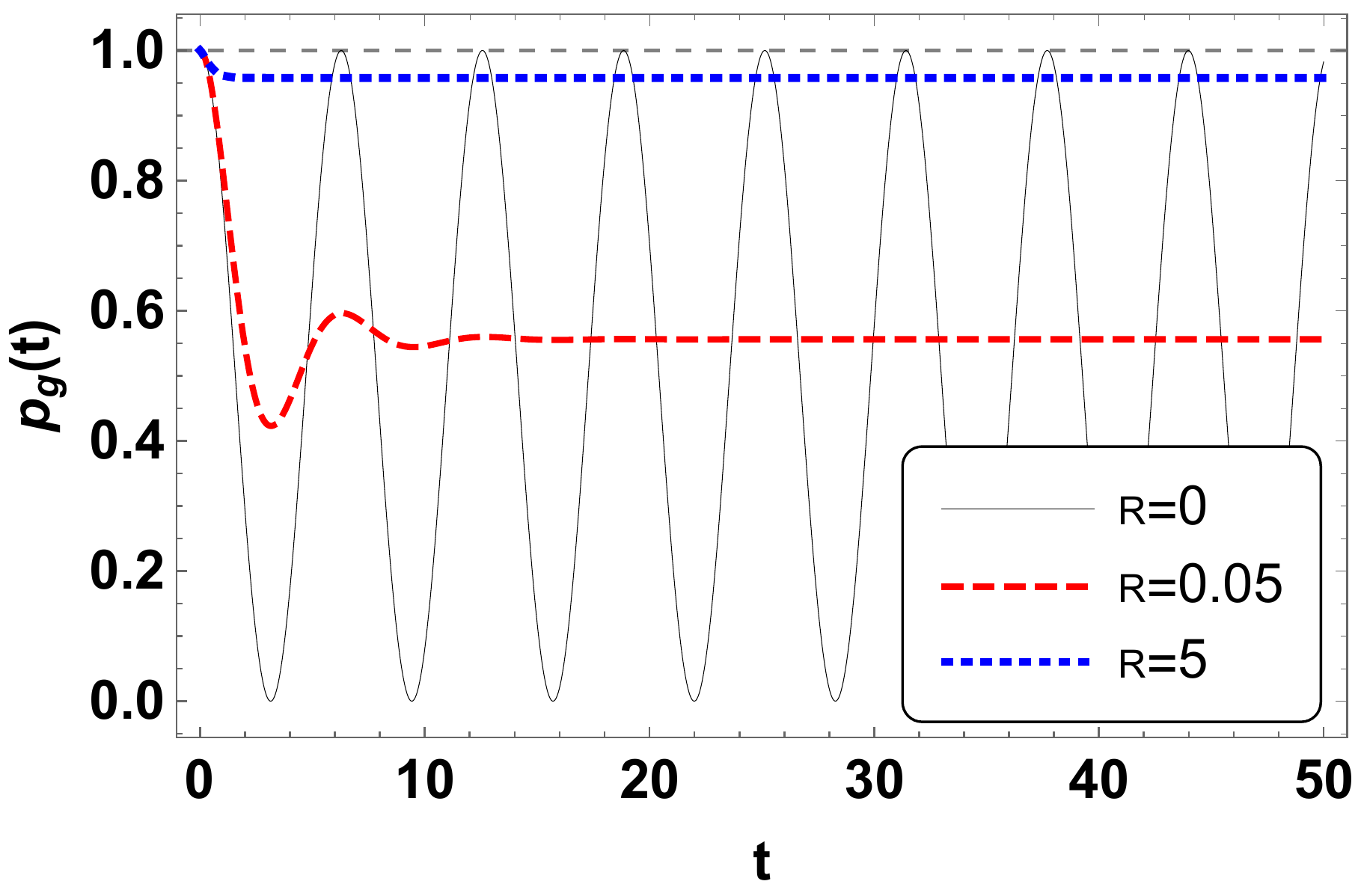}
	 	\caption{(color online). Probability of finding the atom in ground state at time $t$, in the units with $\hbar = k_B = 1$. Here, $R = \gamma_0/\Omega$ is the ratio of the spontaneous emission to the Rabi frequency. With  squeezing parameter $s=0$ and transition frequency $\omega_0 = 0.5$, the values  $R = 0$, $0.05$ and $5$ correspond to  $\mu_s = 1$, $0.9$ (underdamped) and $0.7 i$ (overdamped), respectively.}
	 	\label{fig:Prob}
	 \end{figure}
	 
	 \begin{figure*}
	 	\includegraphics[width=50mm]{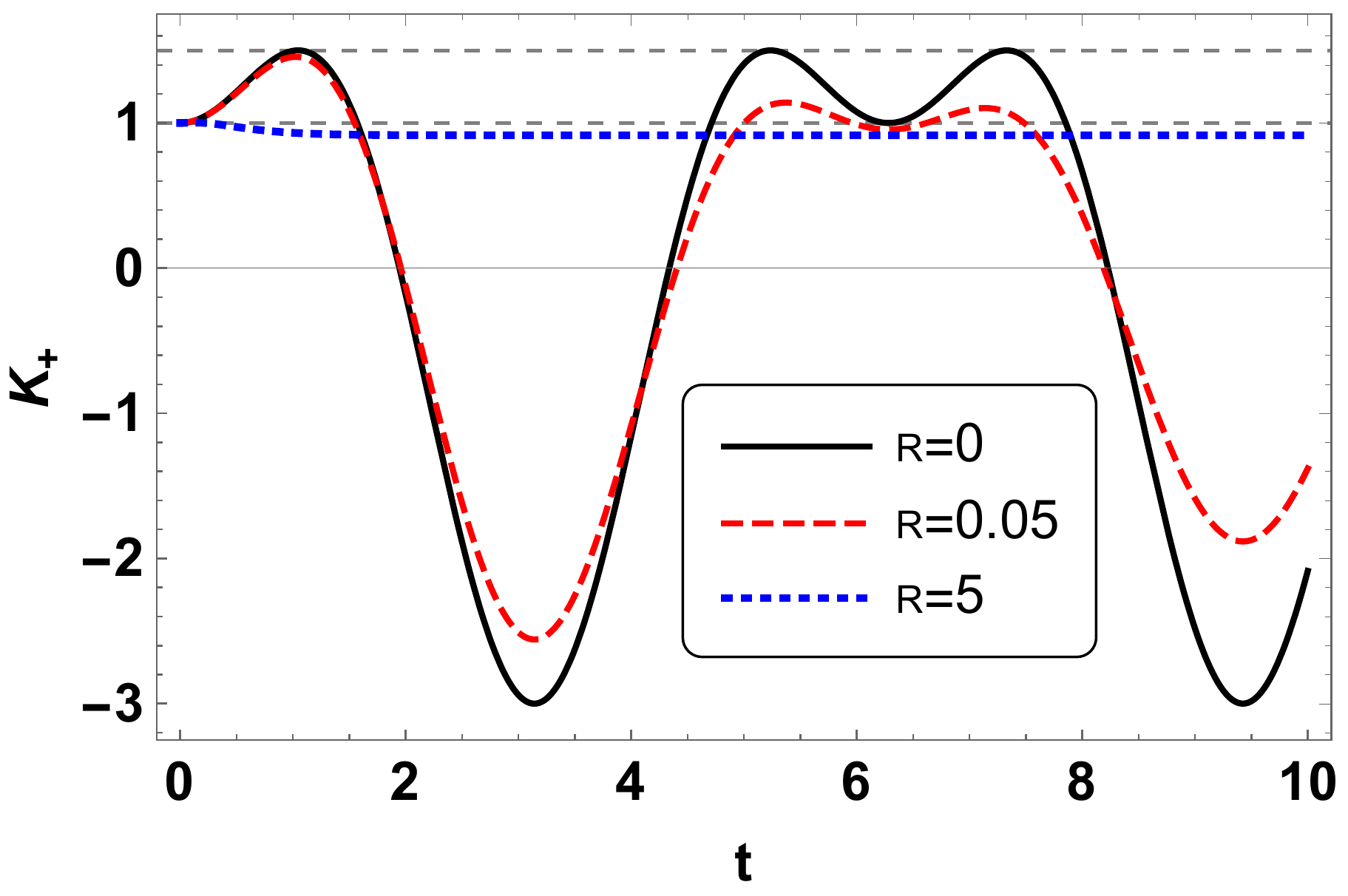}
	 	\includegraphics[width=50mm]{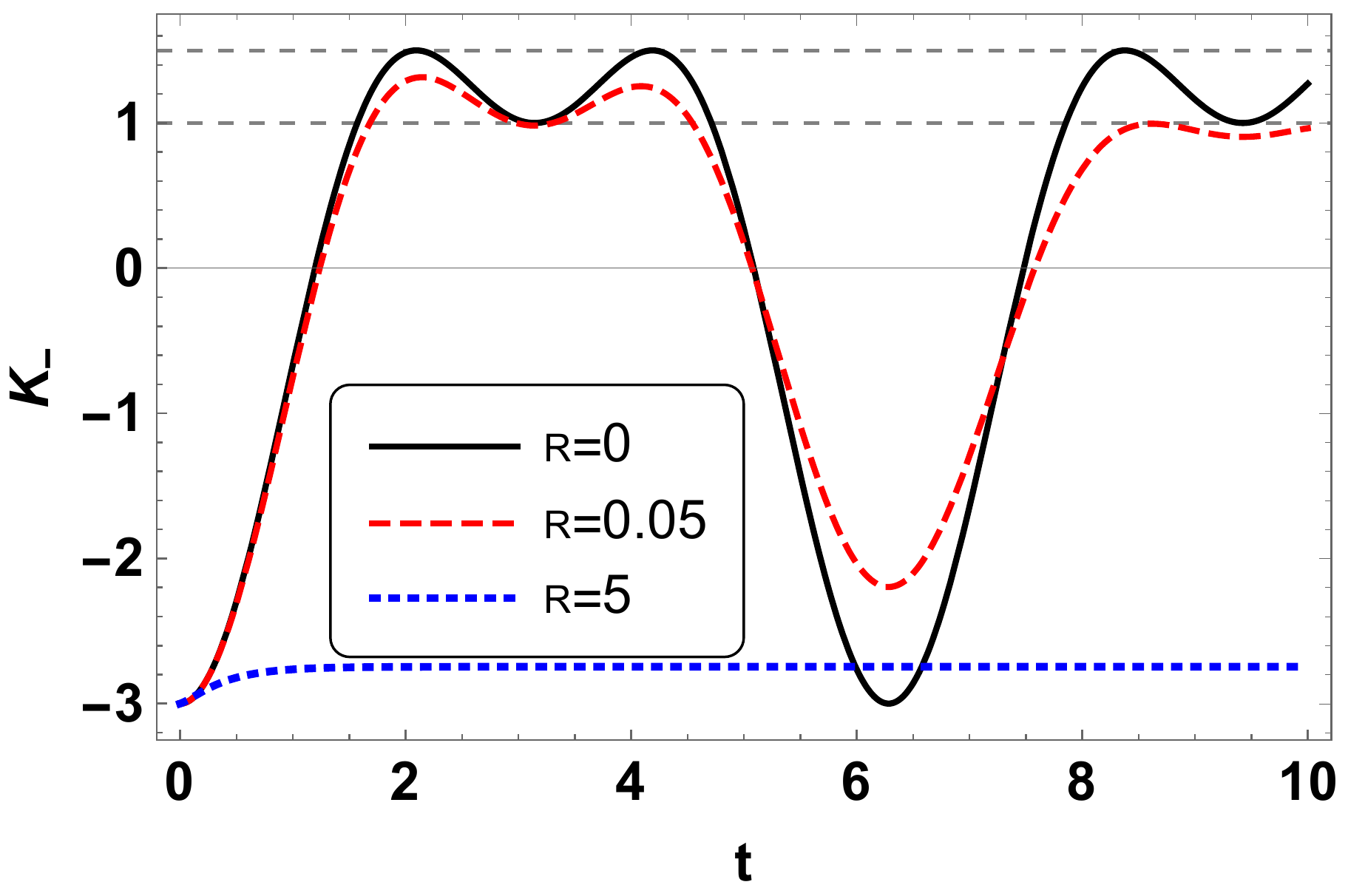}
	 	\includegraphics[width=51mm]{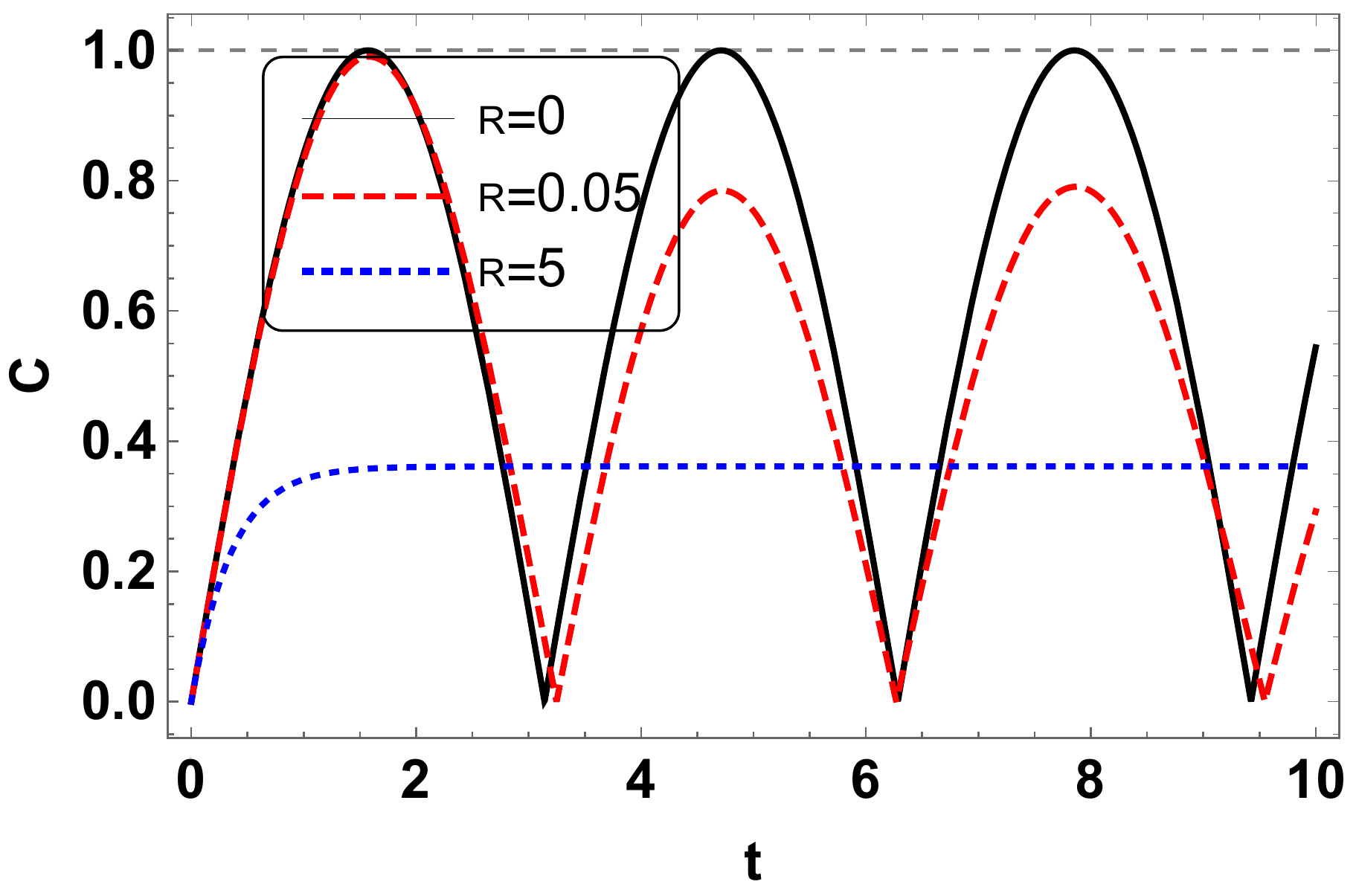}
	 	\caption{(color online). Evolution of the LG parameters $K_{+}$ (left),  $K_{-}$ (middle) and coherence parameter $C$ (right). Here, $\beta = 10$, $\omega_0 = 0.5$, $s=0$, such that $R=0$, $0.05$ and $5$ correspond to $\mu_s = 1$, $0.9$ (underdamped) and $0.7i$ (overdamped) cases, respectively.  The violation of LGtIs  occur predominantly in  underdamped regime such that $K_{\pm}$ reach their quantum bound $3/2$ as $R \rightarrow 0$. The coherence parameter shows exponentially damped oscillations in underdamped regime, while in overdamped case, it monotonically saturates to it stationary value.}
	 	\label{fig:Kplus_Kminus}
	 \end{figure*}
	 
	  Consequently, the stationary population of the excited state $p^s_e =  \frac{1}{2}(1 + \langle \sigma_3 \rangle_s ) = \frac{1}{2} \big[ 1 - \frac{\gamma_0 (\gamma - 2 \gamma_0 M)}{\gamma^2 - 2 \gamma \gamma_0 M + 2 \Omega^2} \big]$. 
	  
	    In the strong driving limit, $\Omega \gg \gamma_s$, we have $p_e^s = 1/2$ and $\langle \sigma_+ \rangle_s = -i\gamma_0/2\Omega$.\par
	 In order to solve the time dependent Bloch equation, Eq. (\ref{eq:BlochEq}), it is convenient to introduce the vector 
	 \begin{equation}
	 \langle \vec{\Sigma}(t) \rangle = \langle \vec{\sigma}(t) \rangle - \langle \vec{\sigma} \rangle_s.
	 \end{equation}
	 This vector satisfies the homogeneous equation
	 \begin{equation}\label{eq:SigmaEq}
	 \frac{d}{dt}  \langle \vec{\Sigma}(t) \rangle = \mathcal{G}  \langle \vec{\Sigma}(t) \rangle.
	 \end{equation}
	  This equation can be easily solved by diagonalizing $\mathcal{G}$, which has the eigenvalues
	 \begin{align}
	 \lambda_1 &= -\frac{\gamma}{2} - \gamma_0 M,\nonumber\\
	 \lambda_{2,3} &= \frac{\gamma_0 M}{2} - \frac{3 \gamma}{4} \pm i \mu_s,
	 \end{align}
	 where,
	 \begin{equation}\label{mus}
	 \mu_s = \sqrt{ \Omega^2 - \Big(\frac{\gamma_s}{4} \Big)^2}\quad {\rm with}~~ \gamma_s = \gamma + 2 \gamma_0 M.
	 \end{equation}
	 
	 Assuming the atom to be initially in the ground state $\rho(0) = \ket{g}\bra{g}$, we have 
	 \begin{equation}
	 \langle \sigma_3(0) \rangle = -1\quad {\rm or}\quad \langle \Sigma_3(0) \rangle = -1- \langle \sigma_3 \rangle_s,
	 \end{equation}
	 and 
	 \begin{equation}
	 \langle \sigma_{\pm}(0) \rangle = 0\quad {\rm or}\quad \langle \Sigma_{\pm} (0) \rangle = - \langle \sigma_{\pm} \rangle_s.
	 \end{equation}
With these initial conditions, the solution of Eq. (\ref{eq:SigmaEq}) is given by
\begin{widetext}
	\small
\begin{equation}
\langle \vec{\Sigma} (t) \rangle  = \begin{pmatrix}
e^{-(\gamma + 2 \gamma_0 M)t/2} \langle \Sigma_1(0) \rangle \\\\
e^{(-3 \gamma + 2 \gamma_0 M)t/4} \Big[ \big( \cos(\mu_st) + \frac{\gamma + 3 \gamma_0 M}{ 4 \mu_s} \sin(\mu_s t) \big) \langle \Sigma_2(0) \rangle + \frac{\Omega}{\mu_s} \sin(\mu_s t) \langle \Sigma_3(0) \rangle \Big]\\\\
e^{(-3 \gamma + 2 \gamma_0 M)t/4} \Big[ \big( 1 - \frac{\gamma_0 M}{2 \mu_s}\big) \cos(\mu_s t) - \frac{\gamma}{4 \mu_s} \sin(\mu_s t) \Big] \langle \Sigma_3(0) \rangle + \frac{i \Omega}{\mu_s} e^{(-3 \gamma + 2 \gamma_0 M)t/4} \sin(\mu_s t) \big[ \langle \Sigma_{+}(0) \rangle - \langle \Sigma_{-}(0) \rangle \big]
\end{pmatrix}.
\end{equation}
	\normalsize
\end{widetext}	
Having obtained the solution, one can calculate the survival probability of the atom being in the ground state $\ket{g}$, as
\begin{equation}\label{pg}
 p_g(t) = \frac{1 - [ \langle \Sigma_3 (t) \rangle + \langle \sigma_3 \rangle_s] }{2}.
 \end{equation}
 Further, the degree of coherence is proportional to the off-diagonal element
 \begin{equation}
 \langle \sigma_+ (t) \rangle =\frac{ \langle \sigma_1 (t) \rangle + i \langle \sigma_2 (t) \rangle}{2} + \langle \sigma_+  \rangle_s.
 \end{equation}
 The dynamics is underdamped or overdamped depending on whether $\mu_s$, defined in Eq. (\ref{mus}), is  real or imaginary.  As a result, in underdamped regime, the probabilities as well as the coherence exhibit exponentially damped oscillations, while in the over damped case, they monotonically approach to their stationary values, Fig. (\ref{fig:Prob}). Throughout this paper, we  work in units with $\hbar = k_B = 1$.

	 \section{Leggett-Garg type inequality for the two level driven system}\label{LGI}
	 Let $\mathcal{E}_{t_j \leftarrow t_i}$ be the map corresponding to the evolution given by Eq. (\ref{eq:master}), such that the system in  state $\rho(t_i)$ at time $t_i$ evolves  to state $\rho(t_j)$ at some later time $t_j>t_i$ 
	 \begin{equation}
	 \rho(t_j) = \mathcal{E}_{t_j \leftarrow t_i}[\rho(t_i)].
	 \end{equation}
	Let at time $t_0$ the system be in the ground state $\ket{g}$. We define the dichotomic  observable $\hat{\mathcal{M}} = | g \rangle \langle g| - |e\rangle \langle e|$. Thus a measurement of this observable leads to  $+1$ or  $-1$ depending to whether the system is in the ground or excited state, respectively, Fig. (\ref{fig:ModelExp}). We introduce the projectors $\Pi^+ =| g \rangle \langle g| $ and $\Pi^- = | e \rangle \langle e|$, such that $O = \Pi^+ - \Pi^-$. Using Eq. (\ref{eq:TTC}), with the notation $t_1-t_0 = t$,  the two time correlation $C(t_0, t_1)$ is 
	\begin{align}\label{C0t}
	C(t_0, t_1) &= \operatorname{Tr}[\Pi^+ \rho(t_0)] \operatorname{Tr} \Big[\Pi^+ \mathcal{E}_{t_1 \leftarrow t_0} \big[ \frac{\Pi^+ \rho(t_0) \Pi^+}{\operatorname{Tr}[\Pi^+ \rho(t_0)]}\big]  \Big] \nonumber\\&- \operatorname{Tr}[\Pi^+ \rho(t_0)] \operatorname{Tr} \Big[\Pi^- \mathcal{E}_{t_1 \leftarrow t_0} \big[ \frac{\Pi^+ \rho(t_0) \Pi^+}{\operatorname{Tr}[\Pi^+ \rho(t_0)]}\big]  \Big] \nonumber \\&- \operatorname{Tr}[\Pi^- \rho(t_0)] \operatorname{Tr} \Big[\Pi^+ \mathcal{E}_{t_1 \leftarrow t_0} \big[ \frac{\Pi^- \rho(t_0) \Pi^-}{\operatorname{Tr}[\Pi^- \rho(t_0)]}\big]  \Big] \nonumber \\&+ \operatorname{Tr}[\Pi^- \rho(t_0)] \operatorname{Tr} \Big[\Pi^- \mathcal{E}_{t_1 \leftarrow t_0} \big[ \frac{\Pi^- \rho(t_0) \Pi^-}{\operatorname{Tr}[\Pi^- \rho(t_0)]}\big]  \Big],\nonumber \\
	           &= p_g (t) - p_e(t) = 2 p_g(t) - 1.
	\end{align}
	Plugging in the expressions of probabilities, we have 
	\begin{equation}
	K_{\pm} = \pm 2 \mathcal{F}(t) - \mathcal{F}(2t) \mp 1.
	\end{equation}
	Here,
	\small
	\begin{equation}
	\mathcal{F}(t) = \mathcal{A} \big[ \mathcal{B} + \mathcal{C} e^{-(3 \gamma - 2 \gamma_0 M)t/4} \cos(\mu_s t) + \mathcal{D} \sin(\mu_s t) \big] - 1,
	\end{equation}
	 with coefficients given by
	\begin{align}
	\mathcal{A} &= \big[ 4 \mu_s (\gamma^2 - 2 \gamma \gamma_0 M + 2 \Omega^2)\big]^{-1},\nonumber\\
	\mathcal{B} &= 4 (\gamma + \gamma_0)(\gamma - 2 \gamma_0 M) \mu_s + 8 \mu_s \Omega^2, \nonumber\\
	\mathcal{C} &= -2(\gamma_0 M - 2\mu_s) \big[ (\gamma-\gamma_0)(\gamma - 2\gamma_0 M) + 2 \Omega^2\big],\nonumber \\
	\mathcal{D} &= - \gamma (\gamma - \gamma_0) (\gamma - 2\gamma_0 M) - 2 (\gamma - 4\gamma_0) \Omega^2.
	\end{align}
	\normalsize
 In the strong driving limit, $\Omega \gg \gamma_s$, the coefficients can be approximated as $\mathcal{A} \approx \Omega^{-3}$, $\mathcal{B} \approx  \mathcal{C} \approx \Omega^3$ and $\mathcal{D} \approx \Omega^2$, such that in this limit, $\mathcal{F}(t) \propto \cos(\Omega t)$ and therefore
 \begin{equation}\label{strongDrivig}
 K_{\pm} \approx \pm 2\cos( \Omega t) - \cos(2 \Omega t).
 \end{equation}

\begin{figure}
	\includegraphics[width=70mm]{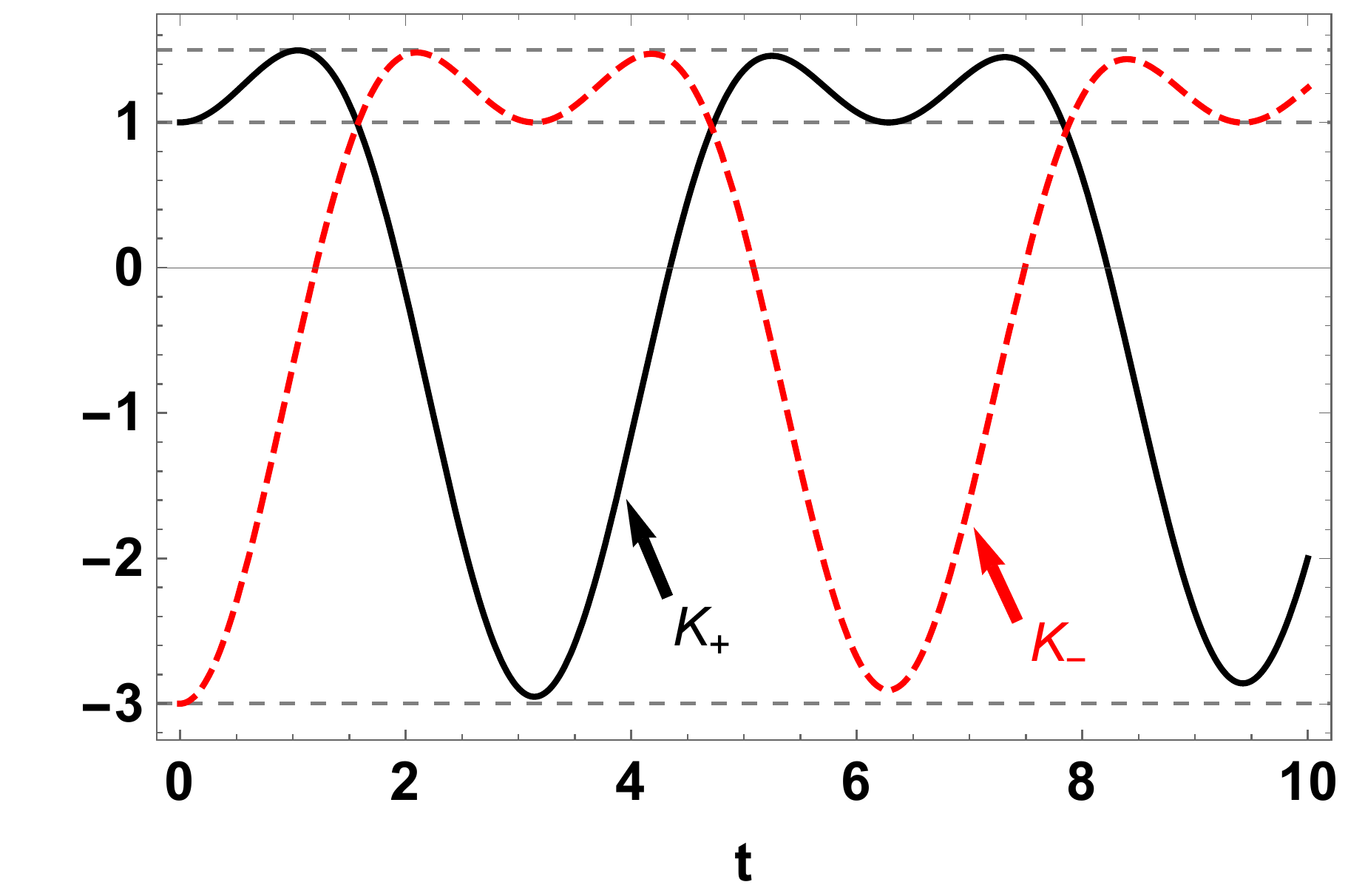}
	\caption{(color online). Complementary behavior of LG parameters $K_{\pm}$ in the strong driving limit. The various parameters used are $\beta = 10$, $\omega_0 =0.5$, $s=0$, $R = 0.005$,   pertaining to the underdamped regime.}
	\label{fig:KplusKminus}
\end{figure}

\textit{Effect of weak measurement}:  The two time correlation function $C(t_0,t)$,  Eq. (\ref{C0t}), was obtained by assuming that the measurements are ideal or projective. However, it would be interesting to see how weak measurements affect the behavior of $C(t_0,t)$ and thereby of the LG parameters $K_{\pm}$. The weak measurements are characterized by invoking a parameter $\xi$ \cite{BuschUnsharp1986,SahaUnsharp}, such that the ideal projectors $\Pi^\pm$ are replaced by the  ``weak projectors" $W^\pm$ defined as
\begin{equation}
W^{\pm} = \Big(\frac{1 \pm \xi}{2} \Big) \Pi^+  + \Big(\frac{1\mp \xi}{2}\Big) \Pi^-.
\end{equation} 
Here, $0 < \xi \le 1$, such that when $\xi = 1$,  $W^\pm$ reduce to the ideal projection operators $\Pi^\pm$. Invoking weak projectors leads to the following form of the two time correlation function becomes $C(t_0,t)|_{weak} = \xi^2 C(t_0,t)$, and consequently
\begin{equation}
K_{\pm}|_{weak} = \xi^2 K_{\pm}.
\end{equation}
Therefore, the maximum violation of LGtI occurs for an ideal projective measurement.
\begin{figure}
	\includegraphics[width=70mm]{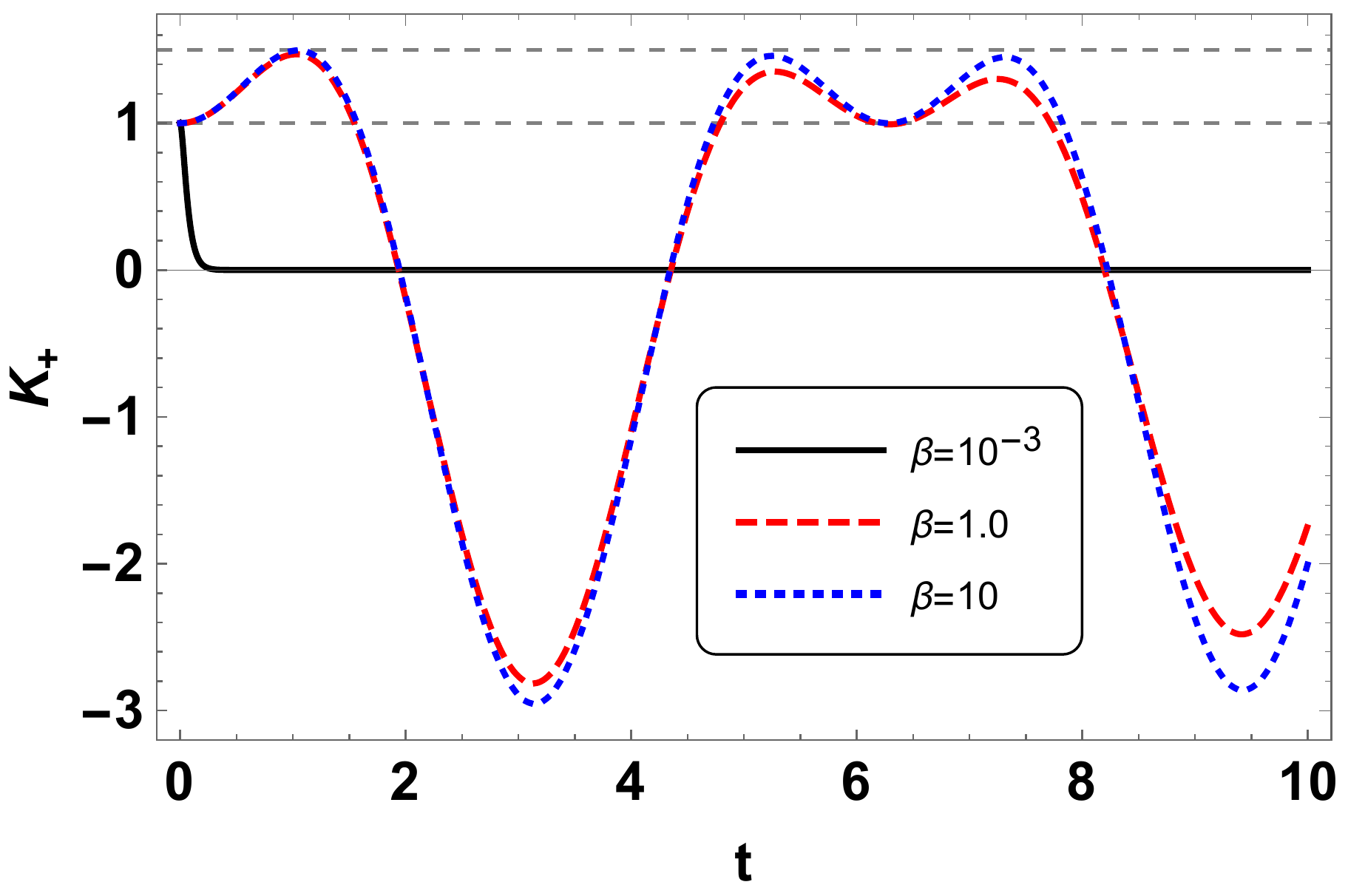}
	\caption{(color online). Temperature dependence of  LG parameter $K_+$. With $\omega_0 = 0.5$, $s=0$ and $R = 0.005$,  the values  $\beta =10$, $1$ and $10^{-3}$ correspond to $\mu_s = 1$, $0.9$ (underdamped) and $4.8i$ (overdamped), respectively.}
	\label{fig:KplusTemp}
\end{figure}
\section{Results and discussion}\label{R&D}
The LGtIs given by inequality (\ref{eq:LGtI}) are studied in the context of a  two level atom with the ground and excited states labelled as $\ket{g}$ and $\ket{e}$, respectively. An external field is driving the transition between the two levels. Further, the atom is allowed to interact with a squeezed thermal bath.   The inequalities thus obtained are in terms of experimentally  relevant parameters. The violation of LGtIs occur predominantly in the underdamped regime which is characterized by the real values of parameter $\mu_s$ defined in Eq. (\ref{mus}), such that

\begin{align}
\Omega &> \frac{\gamma_s}{4} = \gamma_0\frac{(2n+1) + 2  M}{4}~~~ {\rm underdamped},\nonumber\\
\Omega &< \frac{\gamma_s}{4}=\gamma_0\frac{(2n+1) + 2  M}{4}~~~ {\rm overdamped}.
\end{align}

 Figure (\ref{fig:Kplus_Kminus}) depicts the behavior of LG parameters $K_{\pm}$ with respect to time $t$, for different values of the ratio $R=\gamma_0/\Omega$.  The violations of LGtIs are observed mainly in the underdamped regime and fade quickly with the increase in  $R$. In other words, strong driving favors the violation of LGtIs to their maximum quantum bound. The right most panel of the figure shows coherence paramter $C$ \cite{alok2016quantum,bhattacharya2016evolution} which is defined as 
\begin{equation}
C = \sum\limits_{i \ne j} |\rho_{ij}|.
\end{equation}
The extent of  violation of LGtIs can be seen as a signature of the  degree of coherence in the system.

 In the strong driving limit, i.e., $\Omega \gg \gamma_s$, the LG parameters are given by Eq. (\ref{strongDrivig}) and are plotted in   Fig. (\ref{fig:KplusKminus}). The  parameters  $K_+$ and $K_-$ show complementary behavior  in the sense that when one of these parameters does not show a violation, the other does, together covering the entire parameter range.\par
The interaction with the squeezed thermal reservoir leads to enhancement in the transition rate  which is given by $\gamma = \gamma_0 (2n + 1)$, where $\gamma_0$ is the spontaneous emission rate and $\gamma_0 n$ is the squeezed thermal induced emission and absorption rate. The  interactions with the reservoir are expected to decrease the  quantumness  in the system. This feature is depicted in Fig. (\ref{fig:KplusTemp}), where $K_+$ shows enhanced violations for larger values of  the parameter $\beta$ i.e, for smaller temperature. 

\begin{figure}
	\includegraphics[width=70mm]{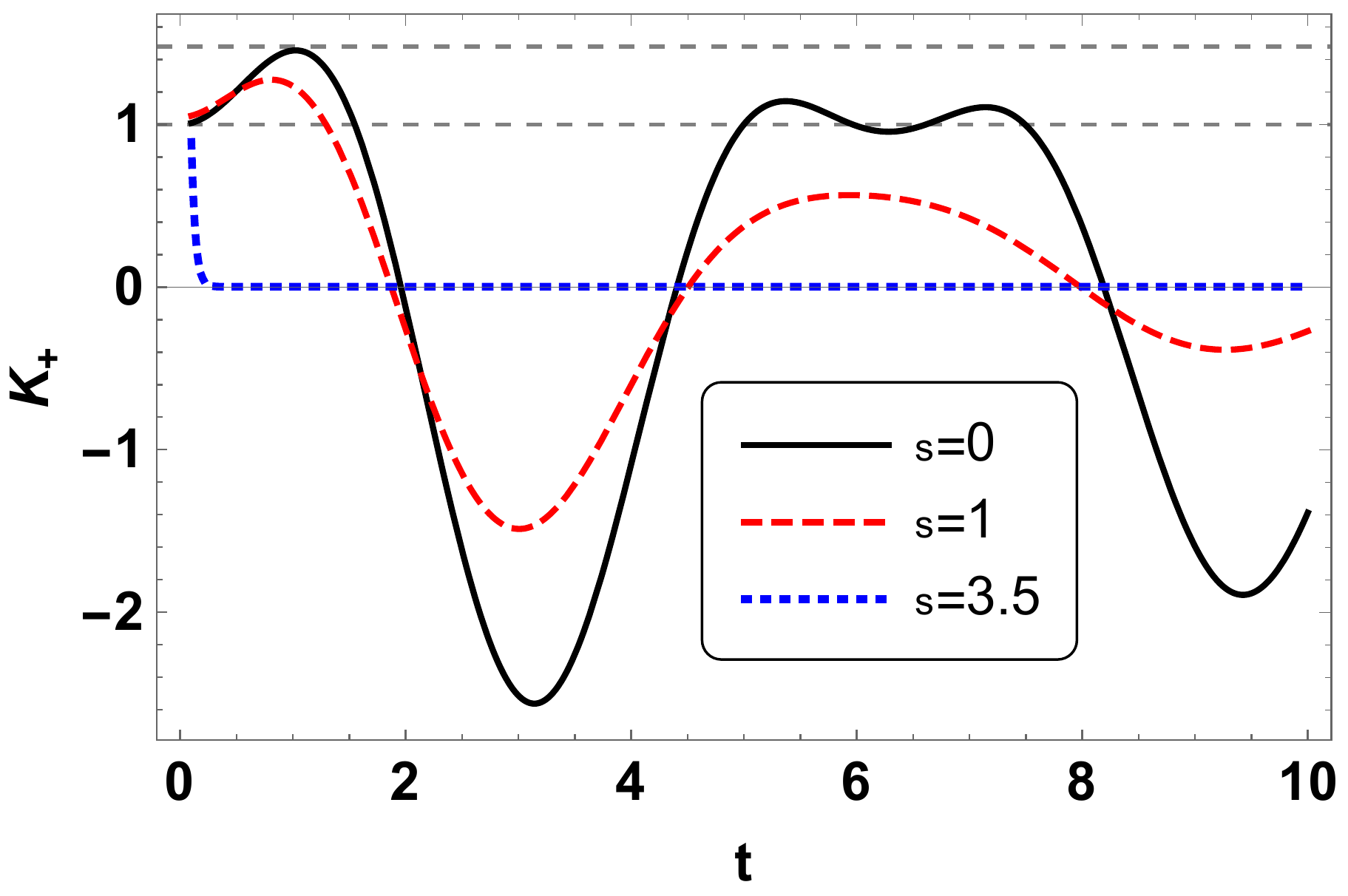}
	\caption{(color online).The LG parameter $K_+$  for different values of the squeezing parameter $s$. Here, $\beta = 100$. $\omega_0 = 0.5$, $R=0.05$. Further, $s=0$, $1$ and $3.5$ correspond to $\mu_s = 1$, $0.9$ (underdamped) and $6.7 i$ (overdamped), respectively.}
	\label{fig:KplusSqueez}
\end{figure}

The squeezing parameter as defined in Eq. (\ref{eq:NandM}), controls the degree of violation of LGtIs, since it affects the total photon distribution. Figure (\ref{fig:KplusSqueez}) exhibits the variation of the LG parameter $K_+$ for different values of squeezing parameter $s$. The increase in $s$ is found to decrease the extent of violation of  LGtI.

\begin{figure}
	\includegraphics[width=85mm]{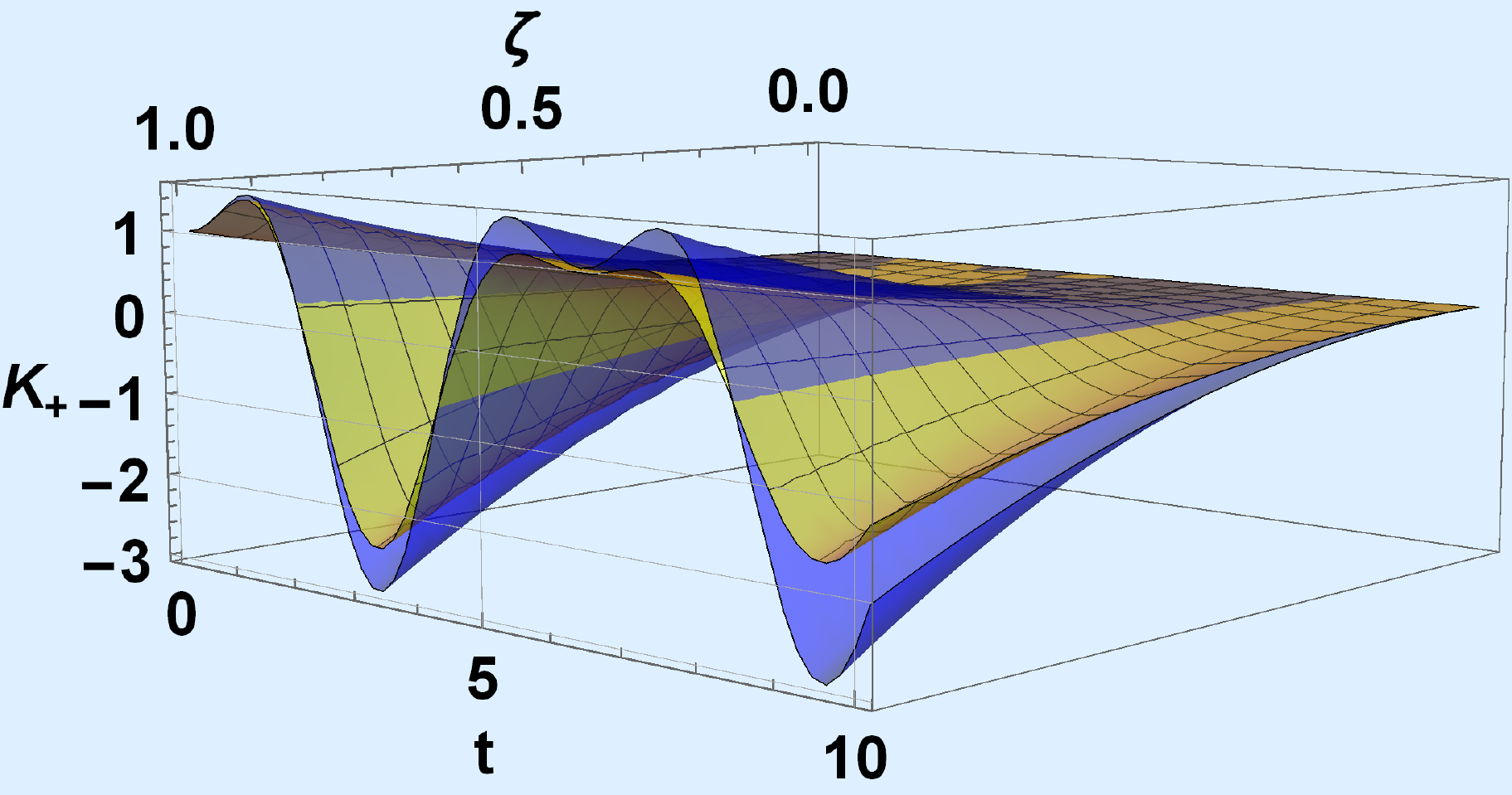}
	\caption{(color online). Variation LG parameter $K_+$  with respect to $t$ and $\xi$. With $\beta = 5$, $\omega_0 = 0.5$ and $s=0$, we have  $R=0$ ($\mu_s \approx 1$) depicted by blue plane-surface,  $R = 0.05$ ($ \mu_s \approx  0.9$) represented by yellow lined-surface. Both these correspond to  underdamped case. 
		% Further,  $R = 5$ ($\mu_s \approx i$) shown by blue surface corresponds to overdamped case.
		 The maximum violation corresponds to $\xi=1$, the ideal projective measurement.}
	\label{fig:KplusUnsharp}
\end{figure}

The effect of weak measurement on the LG parameters is depicted in Fig. (\ref{fig:KplusUnsharp}). The ideal projective measurements are characterized by $\xi = 1$, while  $\xi=0$ corresponds to no measurement. It is clear from the figure that the maximum violation occurs for ideal projective measurements.

\FloatBarrier
\section{Conclusion}\label{Conclusion}
We studied the violation of Leggett-Garg type inequalities in a  driven two level  atom interacting with a squeezed thermal bath. The effect of various experimentally relevant parameters on the violation of the inequality were examined carefully. The violations were seen to be  prominent in the underdamped case. The increase in temperature was found to decrease the degree of violation  as well  as the time over which the violation is sustained. Squeezing the thermal state of the reservoir was also found to reduce the violation of LGtIs. Enhanced violations, reaching to the quantum bound, were witnessed  in the strong driving limit. Further, we studied the effect of the weak  measurements on the extent of violation of LGtI. The weak measurements are characterized by the parameter $\xi$ such that $\xi=0$ ($\xi=1$) corresponds to no measurement (ideal projective measurement). The maximum violation was found to occur for the ideal projective measurements.

\section*{Acknowledgment}
AMJ thanks DST India for J C Bose National Fellowship.
%	\bibliographystyle{apsrev4-1}
%	\bibliography{library} 
%
\end{document}